\documentclass[preprint,12pt]{elsarticle}

\usepackage{amssymb}

\begin{document}

\begin{frontmatter}



\title{Electronic structure of Mott-insulator CaCu$_3$Ti$_4$O$_{12}$: Photoemission and inverse photoemission study}


\author[1]{H. J. Im}
\ead{hojun@hirosaki-u.ac.jp}
\author[1]{M. Iwataki}
\author[1]{S. Yamazaki}
\author[1]{T. Usui}
\author[1]{S. Adachi}
\author[2]{M. Tsunekawa}
\author[1]{T. Watanabe}
\author[1]{K. Takegahara}
\author[3,4]{S. Kimura}
\author[4,5]{M. Matsunami}
\author[6]{H. Sato}
\author[6]{H. Namatame}
\author[6]{M. Taniguchi}

\address[1]{Graduate School of Science and Technology, Hirosaki University, Hirosaki 036-8224, Japan}
\address[2]{Faculty of Education, Shiga University, Otsu 520-0862, Japan}
\address[3]{Graduate School of Frontier Biosciences, Osaka University, Suita 565-0871, Japan}
\address[4]{UVSOR Facility, Institute for Molecular Science, Okazaki 444-8585, Japan}
\address[5]{School of Physical Sciences, The Graduate University for Advanced Studies, Okazaki 444-8585, Japan}
\address[6]{Hiroshima Synchrotron Radiation Center, Hiroshima University, Higashi-Hiroshima 739-0046, Japan}

\begin{abstract}
We have performed the photoemission and inverse photoemission experiments to elucidate the origin of Mott insulating states in A-site ordered perovskite CaCu$_3$Ti$_4$O$_{12}$ (CCTO).
Experimental results have revealed that Cu 3$d$-O 2$p$ hybridized bands, which are located around the Fermi level in the prediction of the local-density approximation (LDA) band calculations, are actually separated into the upper Hubbard band at $\sim$ 1.5 eV and the lower Hubbard band at $\sim$ $-$1.7 eV with a band gap of $\sim$ 1.5-1.8 eV.
We also observed that Cu 3$d$ peak at $\sim$ $-$3.8 eV and Ti 3$d$ peak at $\sim$ 3.8 eV are further away from each other than as indicated in the LDA calculations.
In addition, it is found that the multiplet strucutre around $-$9 eV includes a considerable number of O 2$p$ states.
These observations indicate that the Cu 3$d$ and Ti 3$d$ electrons hybridized with the O 2$p$ states are strongly correlated, which originates in the Mott-insulating states of CCTO.
\end{abstract}


\end{frontmatter}


\section{Introduction}

Physics of Mott-insulator has been an underlying concept to understand the intrigue phenomena in transition metal oxides such as perovskite-type materials \cite{Mott74,Imad98}.
One of the most well-known systems is the high-temperature superconductor with the layered perovskite structure, e.g. La$_{2-x}$Sr$_x$CuO$_4$ and La$_{2-x}$Ba$_x$CuO$_4$ \cite{Bedn86}.
The perovskites LaVO$_3$ and YTiO$_3$ are also typical Mott-insulators with the antiferromagnetic and ferromagnetic ordering, respectively \cite{Maha92,Okim95}.
Generally, Mott-insulating states can be understood within the framework of the Mott-Hubbard model as its name says.
When the on-site repulsive Coulomb energy between the $d$-electrons ($U$) is larger than the bandwidth of the $d$-states ($W$), the Mott-insulating phase appears in contrast to the band insulator where the $d$-shell is fully occupied \cite{Mott74,Imad98}.
However, there exists many sophisticated situations rather than an ideal situation to be explained simply.
For instance, the layered perovskite Ca$_2$RuO$_4$ has shown the orbital selective Metal-insulator transition \cite{Alex99,Anis02,Neup09}.
In Sr$_2$IrO$_4$, it has been reported that Mott insulator develops due to the strong spin-orbit interaction of Ir 5$d$-electrons \cite{Kim08}.
In addition, it has been well known that the physics of the Mott insulating states plays an important role to understand the metallic peroskite CaVO$_3$ and SrVO$_3$ \cite{Mori95,Eguc06}.

Mott-insulating A-site ordered perovskite CaCu$_3$Ti$_4$O$_{12}$ (CCTO) has attracted much attention in both fundamental science and technological application, since the discovery of the extremely high dielectric constant ($\varepsilon$) in 2000 \cite{Subr00}.
However, the mechanism of the high $\varepsilon$ has not been fully understood yet, and it is still debating whether the high $\varepsilon$ is intrinsic or extrinsic \cite{Home01,Zhu07}.
This requires a study to clarify the fundamental properties such as the electronic structure.
Recently, we have successfully observed the band dispersion of CCTO in the angle-resolved photoemission (ARPES) experiments and revealed that CCTO is a Mott-type insulator caused by the strong correlation effects:
the Cu 3$d$-O 2$p$ hybridized bands are shifted toward the lower energy side by $\sim$ 1.7 eV and show band narrowing around $-$2 eV in comparison with the local-density approximation (LDA) band calculations \cite{Im13}.
However, in order to fully understand the mechanism of the Mott insulating states in CCTO, the electronic structures in both the occupied and unoccupied regions should be investigated.
In this paper, we report the electronic structures of CCTO in the occupied and unoccupied regions revealed by photoemission (PES) and inverse photoemission spectroscopy (IPES).

\section{Experimental methods}

The polycrystalline sample of CCTO has been synthesized by the conventional solid state reaction technique using CaCO$_3$, CuO, and TiO$_2$ powders.
The well mixed powders with the stoichiometric weights were calcined at 700 $^{\circ}$C for 12 hours.
The pellets pressed by 40 MPa were sintered at 1050 $^{\circ}$C for 24 hours.
These processes have been repeated several times to obtain the high-quality sample.
Single phase of CCTO was confirmed by the X-ray diffraction pattern.

The PES experiments have been performed at the synchrotron radiation beam line BL5U of UVSOR.
The used photon energies ($h \nu$) are 30, 65, and 90 eV.
The total energy resolution is about 75 meV at $h\nu$ = 65 eV.
The IPES experiments have been carried out at HiSOR, using a stand-alone apparatus with the tunable photon energy mode \cite{Sato98}.
The used kinetic energies of incident electrons ($E_k$) are 40 and 46 eV.
The energy resolution was set to about 650 meV at $E_k$ = 46 eV.
All measurements have been performed at the room-temperature ($T$ $\sim$ 300 K).
The clean surface of sample was prepared by \textit{in situ} fracturing in the ultra-high vacuum of $\sim$ 2 $\times$ 10$^{-8}$ Pa.
The Fermi-level ($E_F$) in PES and IPES measurements was referred to that of Au at 300 K and 10 K, respectively.

\section{Results and Discussion}

\begin{figure}
\begin{center}
\includegraphics[width=60mm,clip]{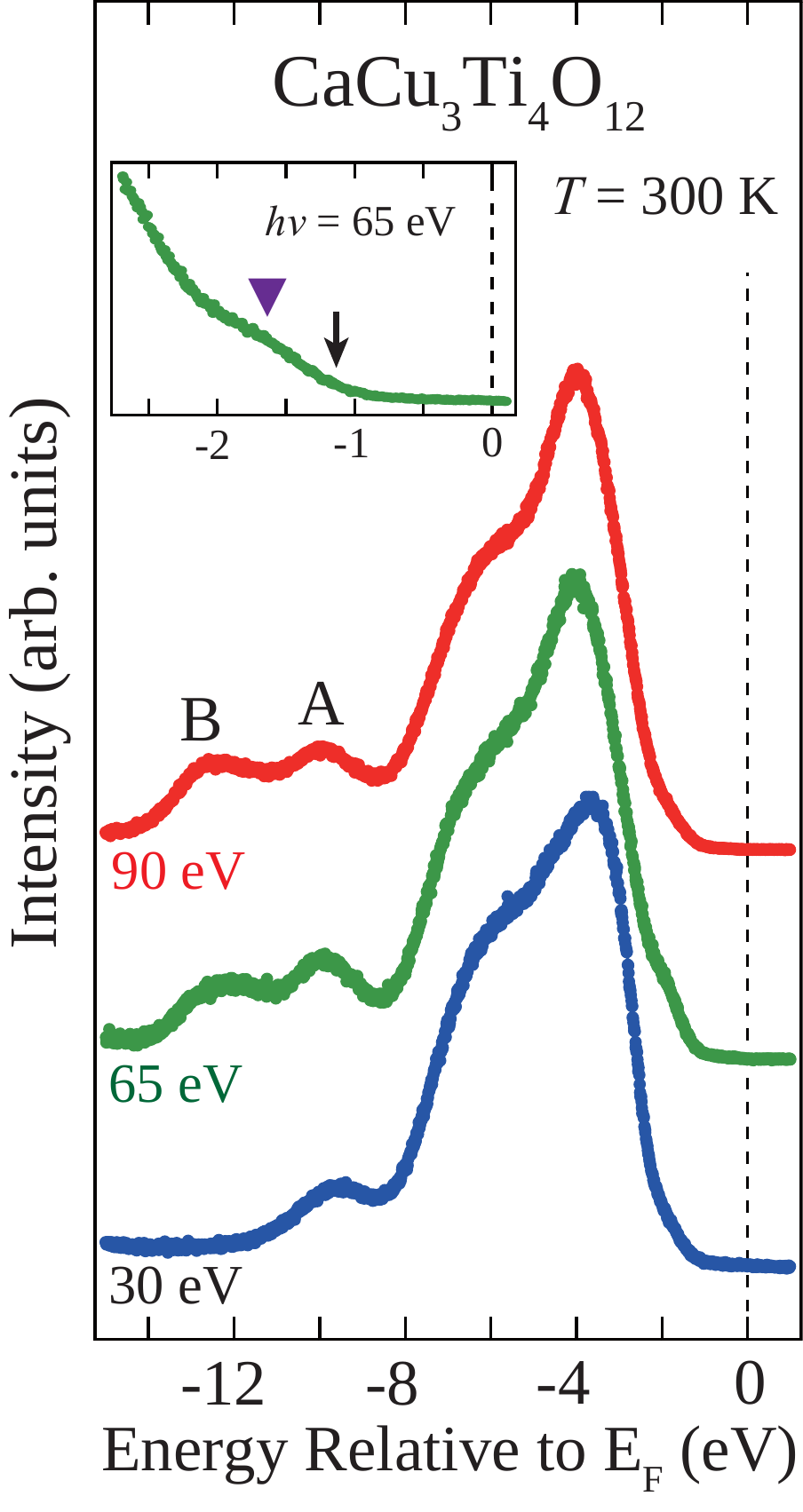}
\caption{\label{fig:Figure1}
  (Color online) PES spectra of CCTO at $h\nu$ = 30, 65, 90 eV and $T$ = 300 K.
  The inset shows the enlarged PES spectrum near $E_F$ at $h\nu$ = 65 eV.
  The solid triangle and arrow indicate the peak position ($\sim$ $-$1.7 eV) and the tail ($\sim$ $-$1.2 eV) of the lower Hubbard band, respectively.
  (In the inset, the $E_F$ of the enlarged PES spectrum was calibrated by shifting toward the higher energy side by 0.2 eV, referring to the previous ARPES results \cite{Im13}.)
  }
\end{center}
\end{figure}

Figure 1 shows the PES spectra of CCTO at $h\nu$ = 30, 65, and 90 eV.
In the region from 0 to $-$8 eV, there are three features in good agreement with the previous ARPES results;
the intense peak around $-$4 eV (mainly Cu 3$d$ states), the shoulder structure around $-$2 eV (mainly Cu 3$d$ and O 2$p$ states) and the hump structure around $-$6 eV (mainly O 2$p$ states) \cite{Im13}.
It is found that the Cu 3$d$ peak position around $-$4 eV is located at $\sim$ 0.2 eV lower energy than that of the previous ARPES experiments \cite{Im13}.
This may come from the charging effects of an insulating sample during the measurement process.
In spite of that, the good consistency of the spectral shape with the ARPES results indicates that the observed PES spectra are revealing the intrinsic electronic structures of CCTO.
The inset shows the enlarged spectrum of the shoulder structure around 2 eV, which was plotted by shifting toward the higher energy side by 0.2 eV to correct $E_F$.
The peak position can be estimated to be $\sim$ $-$1.7 eV, and the tail of the band is also observed at $\sim$ $-$1.2 eV.
In the region from $-$8 to $-$13 eV, there are two peaks, A around $-$9 eV and B around $-$12 eV.
These can be considered as multiplet structures of Cu 3$d$ states hybridized with O 2$p$ states as in CuO which is representative of strongly correlated electrons systems \cite{Im13,Eske90,Kotl06}.
Actually, the valence band of CCTO is very similar to that of CuO except for the position of the intense Cu 3$d$ peak which is located around $-$3 eV in CuO \cite{Eske90}.
(It should be noted that most of the Ti 3$d$ states exist in the unoccupied region and that most of the Ca $sp$-states thinly spread over the whole valence band.)
This means that the electronic structure of Cu 3$d$ states of CCTO is close to that of CuO.
Therefore, the two peaks A and B in CCTO can be understood on the basis of the interpretation of CuO.
The peak A around $-$9 eV is the final states of $d^8$ and $d^{10}\underline{L}^{2}$ ($\underline{L}$ is a hole of ligands), while the peak B around $-$12 eV comes mainly from the $d^8$ final states \cite{Eske90}.
It is recognized that the peak B drastically changes with $h\nu$.
In the photoemission process, it is well known that the photoionization cross section ($\sigma$) depends on the kinetic energy of photoelectrons, which can be changed with $h\nu$ \cite{Yeh85,Hufn95}.
In the $h\nu$ of 30-90 eV, the $\sigma$ of the O 2$p$ orbital rapidly increases with decreasing $h\nu$ compared to that of the Cu 3$d$ orbital \cite{Yeh85}.
The intensity of peak B decreases with decreasing $h\nu$ and seems to disappear at $h\nu$ = 30 eV, while the intensity of peak A does not change with $h\nu$, relatively.
In other words, the relative intensity of peak A to peak B increases with decreasing $h\nu$.
From the above argument, we can conclude that the peak A includes a considerable number of O 2$p$ states while peak B consists mainly of Cu 3$d$ states.
This indicates that the Cu 3$d$-O 2$p$ hybridized bands compose the multiplet structures, which are usually considered to come from atomic states.

\begin{figure}
\begin{center}
\includegraphics[width=70mm,clip]{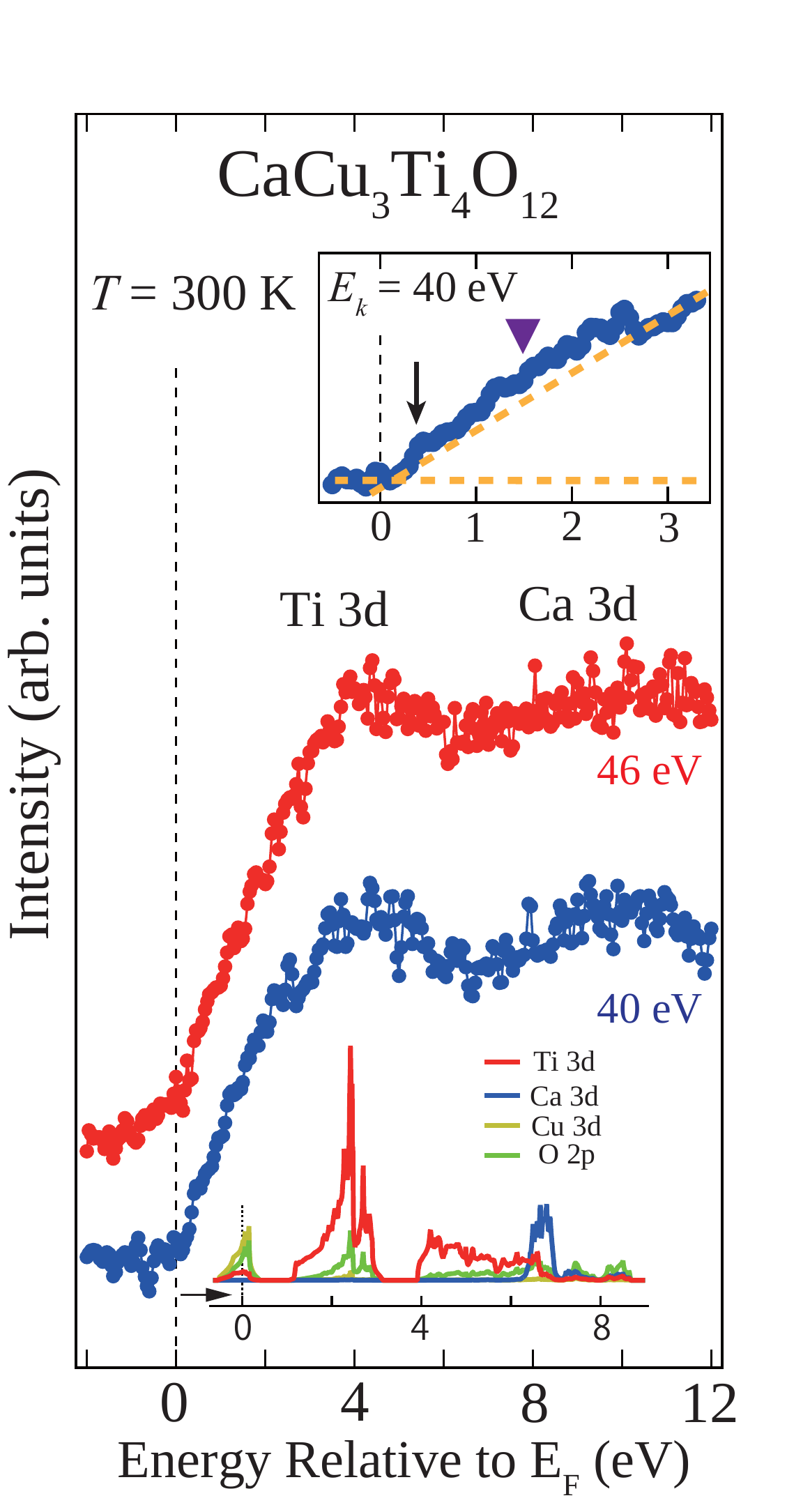}
\caption{\label{fig:Figure2}
  (Color online) IPES spectra of CCTO at $E_k$ = 40, 46 eV and $T$ = 300 K.
  The partial DOS obtained from the LDA calculations is also plotted below the IPES spectra.
  The inset shows the enlarged IPES spectrum near $E_F$ at $E_k$ = 40 eV.
  The solid triangle and arrow indicate the peak position ($\sim$ 1.5 eV) and the tail ($\sim$ 0.3 eV) of the upper Hubbard band, respectively.
  }
\end{center}
\end{figure}

Figure 2 shows the IPES spectra around the Ti 3$p$-3$d$ edge.
The cross section of the Ti 3$d$ states can be largely enhanced around the Ti 3$p$-3$d$ edge via the Fano-resonance between the standard IPES process and the Coster-Kronig Auger process \cite{Imad98,Arit07}.
Additionally, it has been reported that the energies of on- and off-resonance IPES are about 46 eV and 39 eV, respectively \cite{Arit07}.
In the off-resonant IPES spectrum ($E_k$ = 40 eV), we observe three features;
a broad shoulder structure in 0-3 eV, a relatively intense peak at $\sim$ 3.8 eV, and a broad hump structure in 8-12 eV.
The inset of Fig. 2 shows the enlarged spectrum around the broad shoulder structure.
The spectral weight is negligible around $E_F$, indicating an insulating character.
It then begins to increase just above $E_F$, forming a broad shoulder from 0.5 to 2.5 eV.
The peak position can be evaluated from the midpoint of the broad shoulder ($\sim$ 1.5 eV).
When we take into account the energy resolution of 0.65 eV, the tail of the band should exist between 0 to 0.6 eV and its midpoint ($\sim$ 0.3 eV) is indicated by the arrow.
In the on-resonant IPES spectrum ($E_k$ = 46 eV), we find that the peak at $\sim$ 3.8 eV is slightly sharper than that of the off-resonant IPES spectrum.
However, it is difficult to extract the Ti 3$d$ spectrum from the difference between the on- and off-resonant spectra due to the small resonant effects.
In order to discover the reason why the resonance is not effective in CCTO, further study is necessary.
For our purpose here, we identify elements of the spectral weight by comparing the IPES spectra with the band calculations.
In the lower part of Fig.2, we plot the partial density of states (DOS) of Ti 3$d$, Ca 3$d$, Cu 3$d$, and O 2$p$ states obtained from the LDA calculations, shifting it toward the higher energy side by about 1.5 eV.
We can recognize that the shifted partial DOS reproduces the IPES spectra.
The spectral weight of peak at $\sim$ 3.8 eV comes mainly from Ti 3$d$ states, and the broad hump structure in 8-12 eV can be attributed mainly to Ca 3$d$ states.
Moreover, the shoulder structure at $\sim$ 1.5 eV consists of a large number of Cu 3$d$, O 2$p$ states and a small number of Ti 3$d$ states, which are said to be located around $E_F$ in the prediction of the LDA calculations.

\begin{figure}
\begin{center}
\includegraphics[width=75mm,clip]{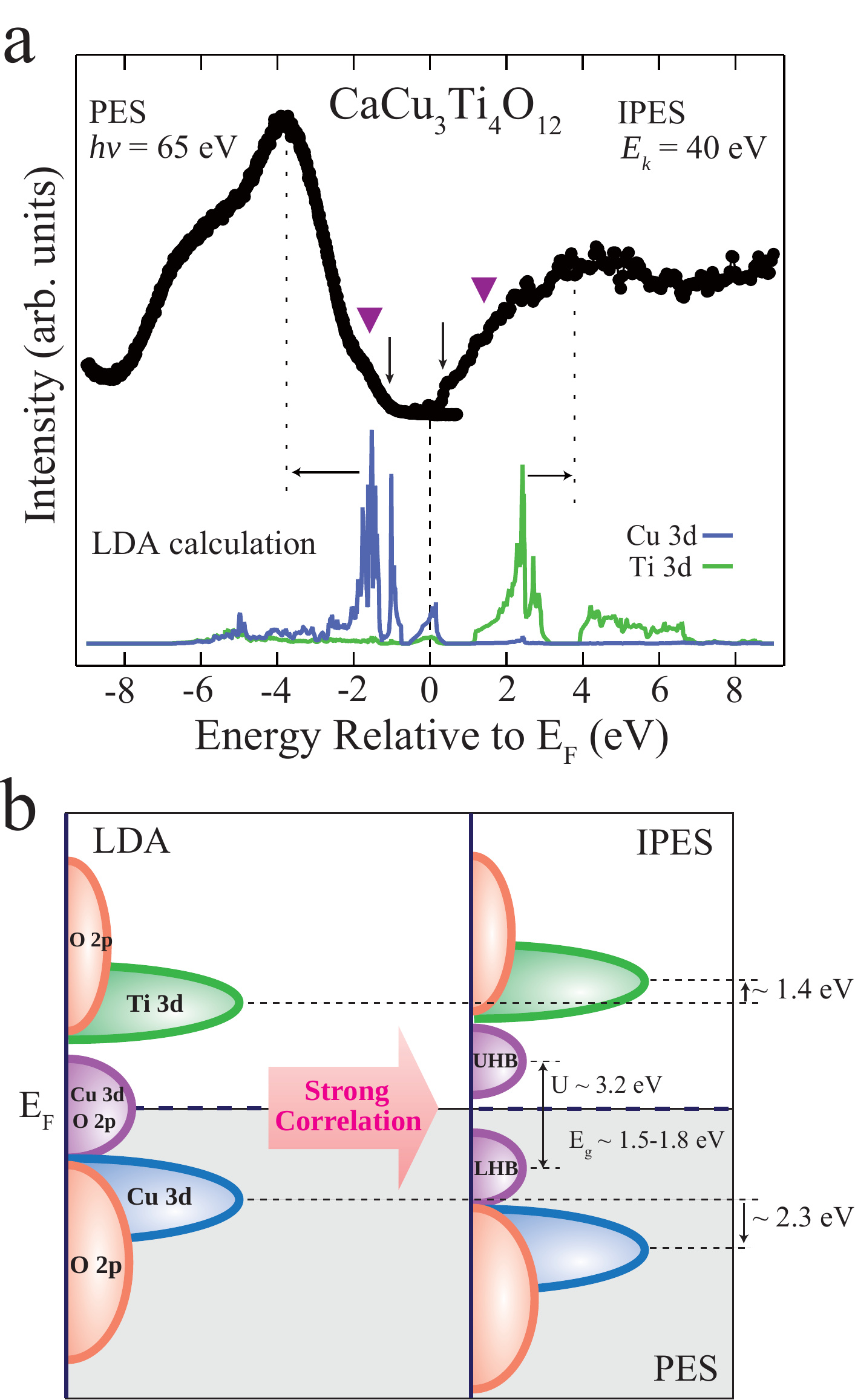}
\caption{\label{fig:Figure3}
  (Color online) (a) The PES and IPES spectra of CCTO obtained at $h\nu$ = 65 eV and $E_k$ = 40 eV, respectively.
  (The $E_F$ of the PES spectrum was calibrated referring to the previous ARPES results \cite{Im13}.)
  The Cu 3$d$ and Ti 3$d$ partial DOS of the LDA calculations are also given in the lower part.
  The solid triangles and arrows on the spectra indicate the peak position and the tail of the Hubbard bands, respectively
  (b) Schematic diagram of the strong correlation effects in CCTO.
  }
\end{center}
\end{figure}

Finally, we would like to discuss the mechanism of the Mott-insulating states in CCTO, considering the strong correlation effects.
In Fig. 3(a), the PES spectrum at $h\nu$ = 65 eV and the IPES spectrum at $E_k$ = 40 eV are plotted together.
For the sake of comparison, the partial DOS of the Cu 3$d$ and Ti 3$d$ states obtained from the LDA calculations has been also presented in the lower part of Fig. 3(a).
It should be noted that the strong correlation effects cannot be fully understood within the framework of the LDA calculations in general and have often been discussed by comparing experimental results and LDA calculations \cite{Imad98,Im13}.
Figure 3(b) is a schematic diagram, which explains how the electronic structures of CCTO are affected by the strong correlation effects.

First, let us discuss the spectra near $E_F$.
The spectral weights are negligible in both PES and IPES experiments showing the insulating phase, while there exists the DOS of the Cu 3$d$ and small Ti 3$d$ states, which are hybridized with the O 2$p$ states, in the LDA calculations.
It is also observed that the shoulder structures at $\sim$ $-$1.7 eV and $\sim$ 1.5 eV are symmetric with respect to $E_F$ as in the other Mott-insulators \cite{Imad98}.
In other words, the DOS near $E_F$ of the LDA calculation is separated into the upper Hubbard band (UHB) at $\sim$ 1.5 eV and the lower Hubbard band (LHB) at $\sim$ $-$1.7 eV \cite{Imad98}.
The energy gap ($E_g$) can be estimated to be $\sim$ 1.5$\pm$0.3 eV from the energy difference between the tails of LHB and UHB.
Moreover, when we take into account the results of the optical experiments, where $E_g$ is larger than 1.5 eV \cite{Home03}, $E_g$ should be $\sim$ 1.5-1.8 eV.
From the above results, $U$ is estimated to be $\sim$ 3.2 eV.
The $W$ of LHB is $\sim$ 1 eV in the PES spectrum, and the $W$ of UHB can be estimated to be $\sim$ 1.5-2 eV considering the experimental resolution in the IPES spectrum.
These values are qualitatively consistent with the simple Mott-Hubbard model, where $E_g = U - W$.

Secondly, let us discuss the spectra in the region from $-$3 to $-$6 eV and from 3 to 6 eV.
In comparison with the LDA calculations, the Cu 3$d$ peak is shifted from $-$1.5 to $-$3.8 eV in the occupied region, and the Ti 3$d$ peak is shifted from 2.4 to 3.8 eV in the unoccupied region.
This seems similar to the behavior of the incoherent states in the simple Mott-Hubbard picture.
However, CCTO looks more complicated.
The observed peak shifts occur far away from $E_F$ and are not directly relevant to the insulating phase.
In addition, the Cu- and Ti-ions are located in different sites (the A'- and B-sites, respectively) and that the O-ions are in between them;
A-site ordered perovskites have a crystal structure of the formula  AA'$_3$B$_4$O$_{12}$ \cite{Subr00,Im13}.
Nevertheless, we can recognize that the shifts of the Cu 3$d$ and Ti 3$d$ peaks appear in pairs and are accompanied by the separation of the Hubbard bands.
For the occupied region, it has been reported in the previous ARPES studies that the strong correlation effects in the Cu 3$d$-O 2$p$ hybridized bands cause the shift of the Cu 3$d$ peak toward the lower energy side \cite{Im13}.
And, the UHB and LHB come from not only the Cu 3$d$-O 2$p$ hybridized bands but also small Ti 3$d$-O 2$p$ hybridized bands as discussed in Fig. 2.
These indicate that the Ti 3$d$ electrons hybridized with O 2$p$ states are also strongly correlated, which causes the shift of the Ti 3$d$ peak toward the higher energy side, as in the case of the Cu 3$d$ electrons.
As a result, the Cu 3$d$ and Ti 3$d$ electrons hybridized with the O 2$p$ states are strongly correlated, which is responsible for the Mott-insulating states and the band shifting in CCTO.

We believe that these results will contribute to the understanding of new-type Mott-insulators and the duality of the strongly correlated $d$-electrons.

\section{Summary}
The mechanism of Mott-insulating behavior of CCTO has been investigated by the PES and IPES experiments.
In the region from $-$2 to 2 eV, UHB and LHB, which consist mainly of Cu 3$d$-O 2$p$ hybridized bands, have been observed at $\sim$ 1.5 and $\sim$ $-$1.7 eV, respectively.
The Cu 3$d$ peak at $\sim$ $-$3.8 eV and Ti 3$d$ peak at 3.8 eV are further away from each other than as predicted in the LDA calculations.
In addition, the strong $h\nu$ dependence of the PES spectrum in the region from $-$8 to $-$13 eV indicates that the multiplet structure A around $-$9 eV includes a considerable number of O 2$p$ states.
These results reveal that the strong correlation effects of the Cu 3$d$ and Ti 3$d$ electrons, which are hybridized with the O 2$p$ states, give rise to the Mott-insulating states of CCTO.







\end{document}